\documentclass[sort&compress, preprint,12pt]{elsarticle}
\usepackage[utf8]{inputenc}
\inputencoding{utf8}
\pdfoutput=1
\usepackage[english]{babel}
\usepackage{hyperref}

\makeatletter
\def\ps@pprintTitle{%
   \let\@oddhead\@empty
   \let\@evenhead\@empty
   \let\@oddfoot\@empty
   \let\@evenfoot\@oddfoot
}
\makeatother

\begin{document}
\def\k{\ensuremath{K_{\mathrm{I}}}}
\begin{frontmatter}

\title{Smoothed finite element method for stress intensity factor estimation: benefits and limitations}

\author{Rokach I.V.\fnref{myfootnote}}
\address{Department of Mechatronics and Mechanical Engineering, Kielce University of Technology, 25-314 Kielce, Poland}
\ead{rokach@tu.kielce.pl}

\begin{abstract}
    The aim of this study was to check how efficient can be smoothed finite element method (FEM) for solution of the linear fracture mechanics problems. Accuracy of stress intensity factor (SIF) computation were investigated using three types of smoothed FEM (namely ES-FEM, NS-FEM and $\alpha$-FEM). Two types of simple benchmark problems (uniaxial tension of the specimens with central and edge cracks) have been considered.  SIF values were calculated by virtual crack extension and modified crack closure integral methods on almost uniform meshes. Influence of the mesh density and type of the method used for SIF determination were investigated too. Conclusions: 1) $\alpha$-FEM has the highest accuracy among all tested methods. 2) In the case of the simplest linear triangles, ES-FEM accuracy is same as in the case of traditional FEM when second order (quadratic) elements are used. 3) NS-FEM always overestimates SIF values. Unfortunately, solutions obtained using this method or $\alpha$-FEM contain unphysical oscillations of displacements and stress near the crack tip.
\end{abstract}

\end{frontmatter}

\section{Introduction}

Calculation of stress intensity factors (SIF) is an important topic of the application of the finite element method (FEM) for solving engineering problems. Naturally, all emerging new types of finite element method are almost immediately tested with regard to their suitability for solving the problems of fracture mechanics. Recently, a new approach to the finite element method called smoothed-FEM (S-FEM) was proposed~\cite{Liu2010}. Within this approach, new methods were developed to change the traditional scheme of using the simplest linear T-elements (triangular for 2D problems and tetrahedral for 3D ones). In classical FEM, utilizing these elements leads to significant jumps in stress and strain on their borders in the high stress gradient zones. To reduce these jumps, various smoothing techniques are used for stress/strain results on the post-processing stage of an analysis. Authors of S-FEM proposed to make gradient smoothing not after solution of the problem, yet before it, on the stiffness matrix calculation stage. Details of various types of smoothed-FEM procedures are presented in the book~\cite{Liu2010}, which summarizes numerous earlier works of their authors. Main features of three of these methods will be presented in the next section briefly.

It is generally known that linear triangular finite elements have low convergence in fracture mechanics problems, poorly approximate the stress field near a crack tip and therefore require a very fine mesh to get acceptable results. Much better perform special singular finite elements used only in the vicinity of crack tip. Previously several types of singular T-elements compatible with plane or solid S-FEM elements have been proposed~\cite{Liu2011, Nguyen2012, Zeng2013}. Although these elements allow computing SIF with high accuracy, they are not free of drawbacks typical for most special finite elements. First, they are based on different shape functions than regular elements. Therefore, additional subroutines have to be added to a finite element program to compute stiffness matrix of a special element. Second, their insertion near crack tip requires either manual changes of the initial mesh to generate additional nodes on selected edges of finite elements or corresponding changes of mesh generation algorithm to insert these nodes automatically.

These shortcomings are usually not important in purely academic research or for highly specialized programs used only for solving fracture mechanics problems. In the universal commercial FE software designed for the mass market, such "alien" elements are usually not used, however. In fact, simplicity of implementation and robustness were crucial reasons of success of so-called quarter-point or Barsoum~\cite{barsoum1975} elements that are commonly used in the most general FE codes for modeling singular stress field near crack tip. 

The aim of this study was to investigate the accuracy of SIF determination using S-FEM on the most coarse and almost uniform grids (without the traditional mesh refinement near the crack tip). Such meshes are typical for topology optimization problems, where speed of calculations is the main priority. Due to this reason, both elements used here (usually -- linear) and overall mesh structure should be as simple as possible. For this type of problem, accuracy of SIF calculation is not a priority. The aim of analysis here is to compare multiple SIF values obtained for different possible cracks situated in various locations in the structure. Even noticeable yet systematic error is acceptable, provided that this error is similar for all analyzed cases. Lately, detailed calculations can be done for a few selected crack configurations to obtain more accurate results.

Another type of optimization problems where this method can be used is the trend track analysis prepared by the industrial CAD-integrated FEM programs, where a relatively high level of calculation error is also compensated by its systematic character.

Approximate SIF calculation on relatively coarse uniform meshes is also used in modeling of multiple rock cracking~\cite{Fu2012}. In such a problem considerable natural variation of strength parameters of the rock is an additional factor that limits the accuracy of these calculations. 

The above reasons have caused the following limitations of this study:
\begin{enumerate}
    \item Only $\k$ determination were considered, as the most important for practice case
    \item All calculations were done using uniform meshes without commonly used increasing mesh density near the crack tip
    \item In all kinds of S-FEM only the simplest linear triangular elements were used
    \item Only the most coarse meshes (with a few elements along the crack length) were considered.
\end{enumerate}

\section{Theory}
\begin{figure}%
    \centering
    \includegraphics{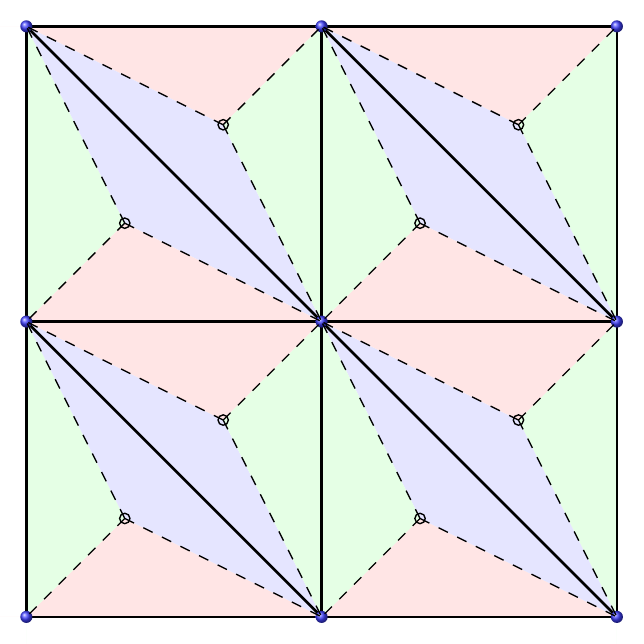}
    \caption{Scheme of ES-FEM smoothed domains build on initial uniform T3-mesh.}%
    \label{fig:es_fem}%
\end{figure}

\noindent As mentioned above, the main innovation used in S-FEM is the preliminary smoothing or averaging of derivatives of the shape functions of several adjacent finite elements prior to calculation of their stiffness matrices. Although this technique can be used for all types of finite elements, it was shown to be the most effective and easy to implement for the linear T-elements~\cite{Liu2010}.  Smoothing is performed in special areas called smoothing domains. Network of the smoothing domains usually is build on the initial mesh used by traditional FEM. Different types of S-FEM use different types of smoothing domains.  

In the case of edge-smoothed FEM (ES-FEM) averaging for a 2D mesh is performed in the quadrilateral areas adjacent to the common edge of two elements~\cite{es}. Their boundaries are shown by dashed lines in Fig.~\ref{fig:es_fem}. In the triangular smoothed domains adjacent to the boundary of the model no smoothing is performed. For the same T-elements mesh, ES-FEM is reported to be very robust and more accurate than traditional FEM~\cite{es,Liu2010}.

\begin{figure}%
    \centering
    \includegraphics{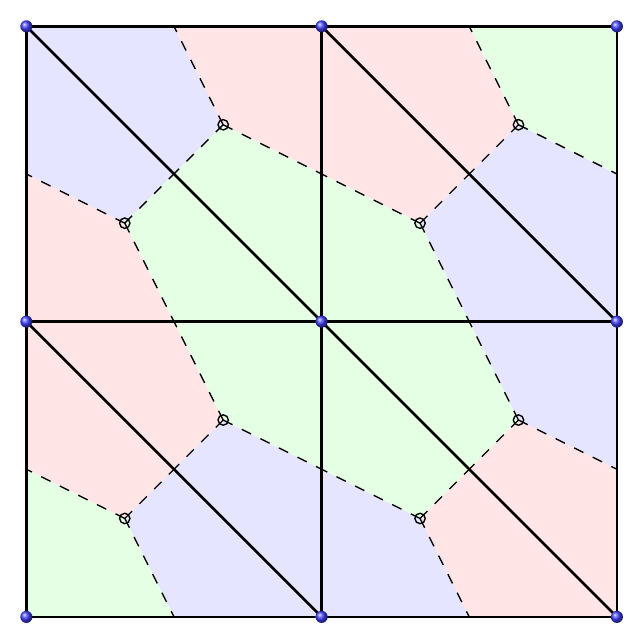}%
    \caption{Scheme of ES-FEM smoothed domains build on initial uniform T3-mesh.}%
    \label{fig:ns_fem}%
\end{figure}

In node-smoothed FEM (NS-FEM)~\cite{ns,Liu2010}, smoothed domains consist of fragments of elements that have a common node of the initial mesh (Fig.~\ref{fig:ns_fem}). An important feature of this type of S-FEM is that it leads to the FE model with reduced total stiffness, i.e. with overestimated total potential energy for force-controlled loading. In contrast, classical FEM always artificially increases model stiffness and therefore underestimates its total potential energy for the same type of load. 

\begin{figure}%
    \centering\includegraphics{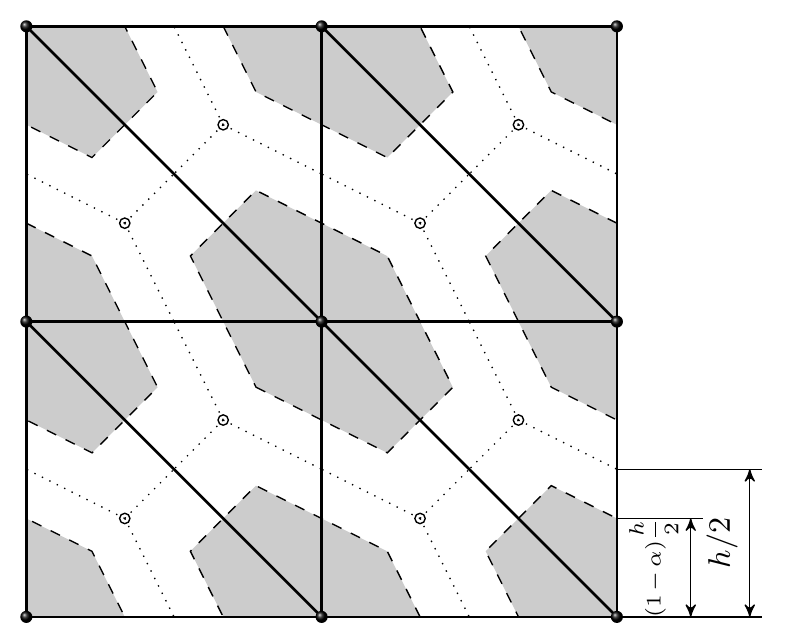}%
    \caption{Scheme of $\alpha$-FEM smoothed domains build on initial uniform T3-mesh. Reduced NS-FEM domains are shadowed.}%
    \label{fig:alpha}%
\end{figure}

There are several ways to utilize this feature. First, it allows obtaining a dual estimation of the results by solving the same problem twice: by traditional FEM and by NS-FEM. Second, stiffness matrix of an element can be calculated partly by NS-FEM using reduced smoothed domain, partially -- by conventional FEM for the rest of the element volume (means for the Y-shaped region in the case of T3 elements in Fig.~\ref{fig:alpha}). Relation between volumes of corresponding parts of an element is controlled by dimensionless parameter $\alpha$.  By selecting the optimal value of this parameter one have an opportunity to get a model with the total potential energy close to the exact value. This type of S-FEM is called $\alpha$-FEM~[\cite{alpha,Liu2010} and definitely is very attractive from the point of view of fracture mechanics due to the close relation existing between SIFs and energy parameters such as an energy release rate. 

Value of parameter $\alpha$ is shown to be mesh and problem dependent~\cite{alpha, Liu2010}. Although a procedure of determination of the optimal $\alpha$ value suggested in Ref.~\cite{Liu2010} is simple and straightforward, it is quite time-consuming. For general 2D problems $\alpha$ values in the range $0.45\ldots0.65$ was shown to be close to the optimal one~\cite{Liu2010}. Calculations performed within this study have shown that for the linear fracture mechanics problems considered in this article corresponding range should be substituted by $\alpha\approx 0.38\ldots0.5$ (details will be published elsewhere). For simplicity, only results for the optimal $\alpha$ value for each problem considered are presented in this work.

\section{Calculations}

\begin{figure}%
    \includegraphics{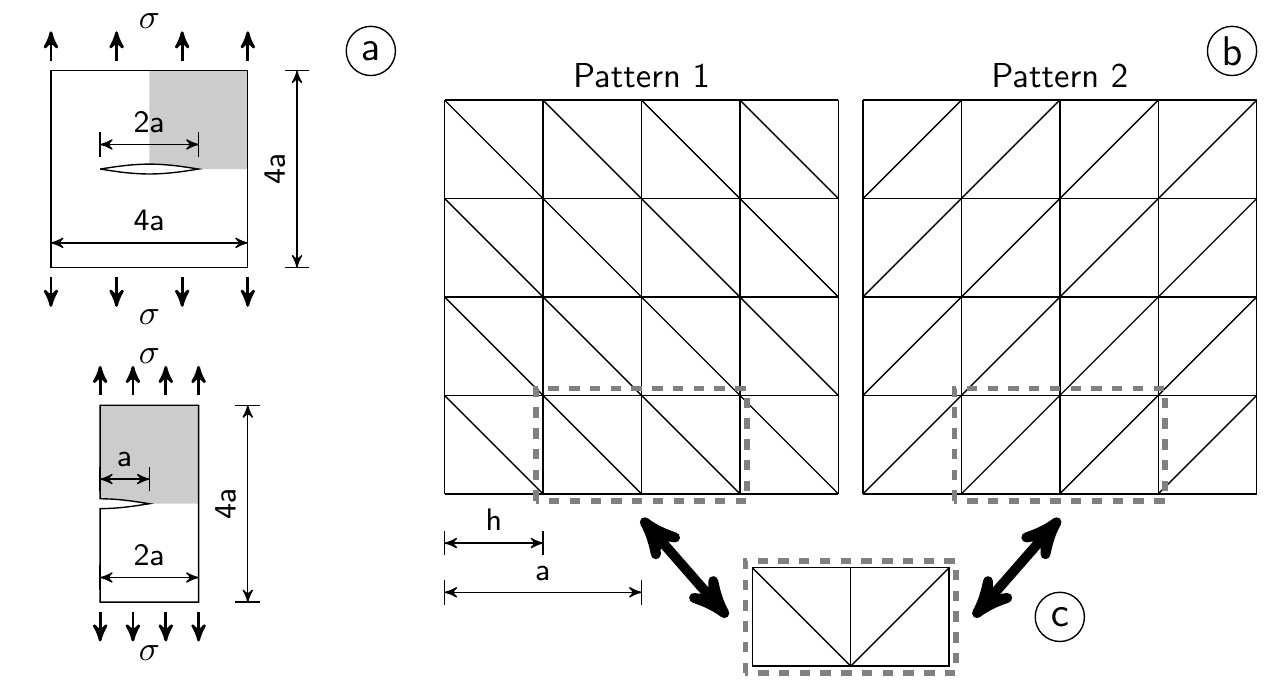}%
    \caption{Schemes of the test problems (a),  typical meshes used in calculations (b) and local modification of the mesh near the crack tip (c).}%
    \label{fig:specimens}%
\end{figure}

\noindent In this paper, three types of S-FEM and traditional FEM were used to solve two test problems, namely uniaxial tension of specimens with a central (CEN) or a side-edge (SEN) cracks (Fig.~\ref{fig:specimens}a). Specimen dimensions were: $2a\times4a$ for the SEN specimen, $4a\times4a$ - for the CEN one, where $a$ is the crack length or the half crack length, correspondingly. Due to symmetry, only a quarter of the CEN specimen and a half of the SEN one were considered (corresponding model parts are shadowed in Fig.~\ref{fig:specimens}a).

In both cases, problems were solved for plane stress conditions using uniform meshes. The following element types have been used:
\begin{enumerate}
    \item The simplest T3 triangles for the smooth and traditional FEM
    \item Polylinear 4-node quads Q4 (full $2\times2$ numerical integration, with and without incompatible modes) for traditional FEM only
    \item Second order 6-node triangles T6 also for conventional FEM only. Node shifting for elements near the crack tip have not been used in this case.
    \item  Second order 8-node quads Q8 (conventional FEM, full $3\times3$ and reduced $2\times2$ numerical integration, quarter-point elements near the crack tip). These elements were used only for determination of the reference solutions (displacements on the crack  surface) of the benchmark problems using very fine mesh.
\end{enumerate}

For all triangular elements two global mesh patterns (Fig.~\ref{fig:specimens}b) have been considered. Additionally, in both cases, small local mesh modification near crack tip was used to increase number of elements in this zone and to investigate the influence of this modification on accuracy of SIF determination (Fig.~\ref{fig:specimens}c).

For all meshes, relation $h/a$ was used as the measure of the mesh density, where $h$ is the distance between nodes on the crack surface which is equal to the element side length for the linear elements and to the half of this length for the quadratic one. Mesh density were changed from $h/a=1/2$ to $h/a=1/32$.

The following methods of \k\ calculation were used:
\begin{enumerate}
    \item Virtual crack extension method (VCE) 
    \item Modified crack closure integral method (MCCI). 
\end{enumerate}

\begin{figure}%
    \includegraphics{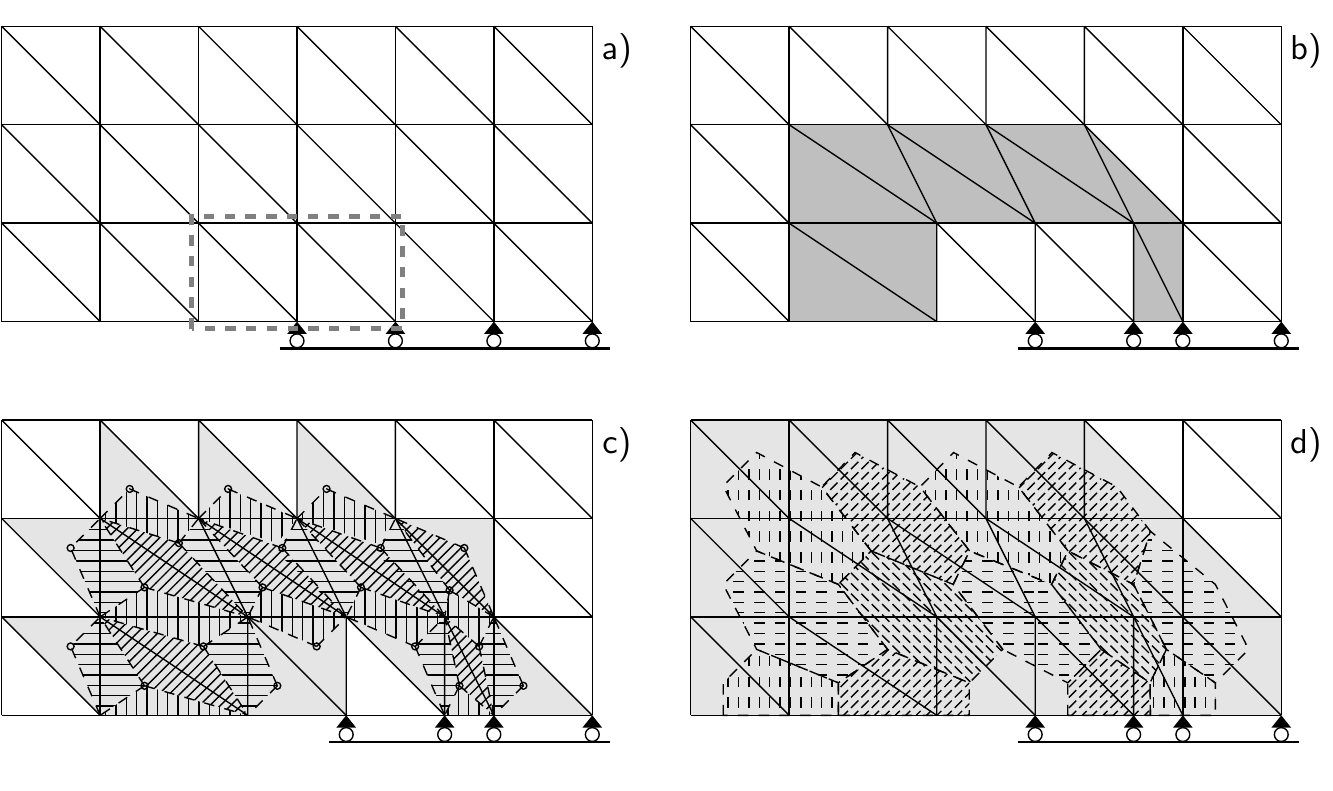}%
    \caption{Scheme of the VCE method: initial mesh with group of 4 elements to be shifted(a), elements distorted due to shifting in traditional FEM (b), elements affected by the same shifting in ES-FEM (c) and NS-FEM (d). }%
    \label{fig:vce_scheme}%
\end{figure}

Calculations have been performed using Matlab scripts created by the author for all T3-elements meshes and by commercial FEM program ADINA 9.0~\cite{adina} for Q8, Q4 and T6 element meshes. Classic form of VCE method was used, namely energy release rate was calculated by finite difference approximation of derivative of the model stiffness matrix~\cite{frac}. When traditional FEM is used, this technique assumes that several (four in Fig.~\ref{fig:vce_scheme}a) FE near the crack tip are shifted to simulate small crack extension. Due to this reason nearby elements (shadowed in Fig.~\ref{fig:vce_scheme}b) are distorted and only their stiffness matrices should be recalculated to estimate the value of derivative of the global stiffness matrix.

SIF values obtained using this method was practically independent on the virtual crack extension value if it was less than $0.01h$. 

In fact, accuracy of the results obtained by VCE and MCCI methods was essentially the same. Difference between them never exceeded 2\%. Complexity of practical implementation of these methods and level of reproducibility of their results were quite different, however.

In classic form of VCE method, partial stiffness matrix of the FE model used for energy release rate calculation should be determined only for those finite elements that are distorted during virtual crack tip shift. In traditional FEM, selection of these elements is quite straightforward. In S-FEM, however, stiffness matrix is calculated for integration domains instead for elements. Due to this reason, in order to use VCE method, one needs to find all distorted domains first and select all elements containing domains after that. Corresponding algorithms are different for each type of S-FEM (see Fig.~\ref{fig:vce_scheme}c and d), complex and time-consuming even for the simplest 2D models used in this study. As usual in VCE method, calculated energy release rate value depends slightly on the size of the shifted zone.  

\begin{figure}%
    \centering
    \includegraphics{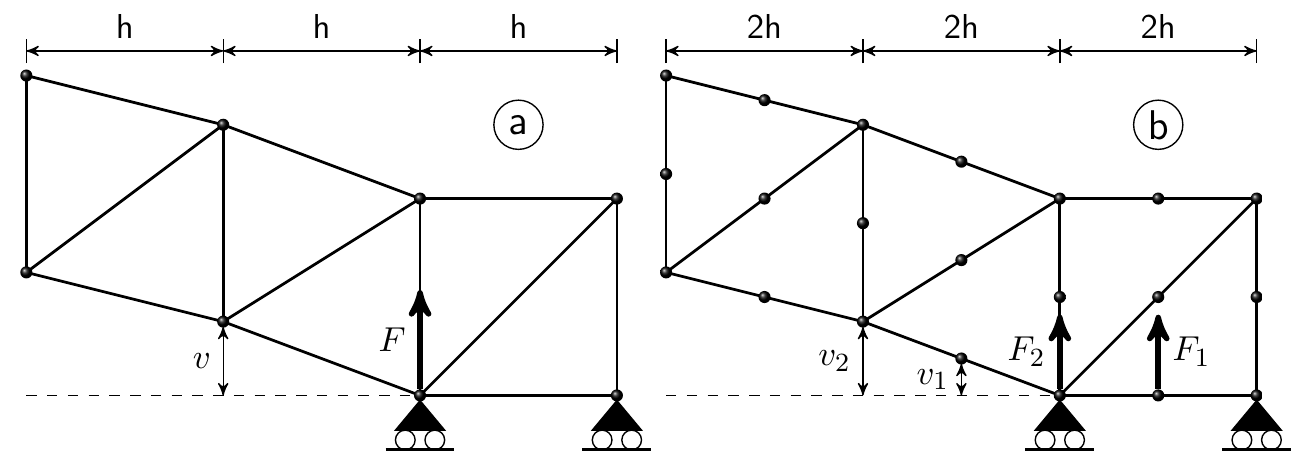}%
    \caption{Quantities used for $G_{\mathrm{I}}$ determination in the MCCI method for linear T3 (a) and quadratic T6 (b) elements}%
    \label{fig:mcci}%
\end{figure}

For linear elements, MCCI method~\cite{frac, mcci} was used in the simplest form, where energy release rate $G_{\mathrm{I}}$ is calculated as 
\begin{equation}
    G_{\mathrm{I}} = \frac{Fv}{h},     
    \label{eq:mcci1}%
\end{equation}
\noindent where $F$ is the nodal force value in the crack tip, $v$ is the half of the crack opening displacement in the closest node to the crack tip (Fig.~\ref{fig:mcci}a). For quadratic elements, the same quantity was calculated as
\begin{equation}
    G_{\mathrm{I}} = \frac{1}{2h}(F_1 v_1 + F_2 v_2),     
    \label{eq:mcciq}%
\end{equation}
\noindent where $F_1$ and $F_2$ are the nodal forces, $v_1$ and $v_2$ are the nodal displacements definied in Fig.~\ref{fig:mcci}b.

Contrary to VCE method, calculations for MCCI one are extremely simple, independent of the type of FEM used (means traditional or smoothed) and give unique result for a particular mesh (there is no virtual shift value and/or distorted domain size dependence). Thus, reproducibility of the results obtained by MCCI method is higher than by VCE one. Due to these reasons, only MCCI results will be presented below.

\section{Results and discussion}

SIF calculation for CEN and SEN specimens in tension are the simplest and most popular benchmarks for linear fracture mechanics problems. In this couple CEN specimen is considered as a 'good cop' (as a rule, all methods of SIF calculation give good results for this problem). Contrary,  SEN specimen is a 'bad cop' because accuracy of SIF obtained for this type of problem is usually the lowest. Due to this reason, SEN specimen related results will be presented below first and investigated more thoroughly.  

For both problems considered, influence of mesh pattern on accuracy of SIFs was quite low. Usually pattern 2 mesh (see Fig.~\ref{fig:specimens}b) gave a bit more accurate results than pattern 1, yet the difference never exceeds 2\%. Additional mesh modification of  near the crack tip (Fig.~\ref{fig:specimens}c) improved accuracy of the results only about 1\%. Due to these reasons, results for unmodified pattern 2 meshes are presented below. 

\begin{figure}%
    \centering
    \includegraphics{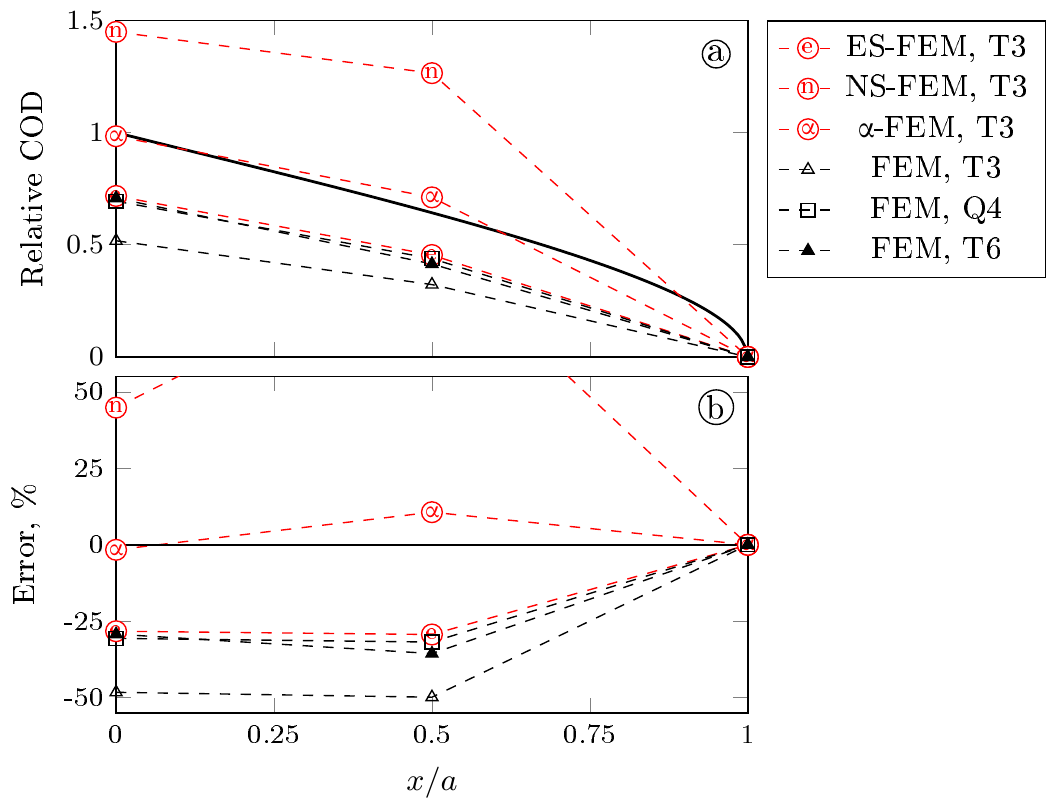}%
    \caption{Comparison of relative crack opening displacements (COD) for the half of the SEN specimen (a) and their accuracy (b) calculated using different methods. Mesh density -- two linear or one quadratic elements per crack length}%
    \label{fig:sen_4_displ_nel4}%
\end{figure}

In Fig.~\ref{fig:sen_4_displ_nel4} displacements on the crack surface of the SEN specimen, obtained using the most coarse meshes ($h/a=1/2$, means having only two linear or one quadratic element on a crack surface) are presented. All results were normalized by the similar displacement values obtained using very fine mesh of Q8 elements. Corresponding shape of the crack surface is shown by solid line in Fig.~\ref{fig:sen_4_displ_nel4}a. It is easy to see that traditional FEM strongly (about 50\%) underestimate COD values. Results obtained by ES-FEM are nearly twice more accurate and are a bit better than the results obtained using formally more precise Q4 and T6 elements. 

Contrary to the edge-smoothed and traditional FEM, NS-FEM essentially (nearly twice in the node closest to the crack tip) overestimates COD values. This overestimation, however, being combined with underestimation of traditional FEM allows $\alpha$-FEM results (obtained for $\alpha=0.42$) to be the most precise. This $\alpha$ value were used for all SEN-calculations in this study. 

\begin{figure}%
    \centering
    \includegraphics{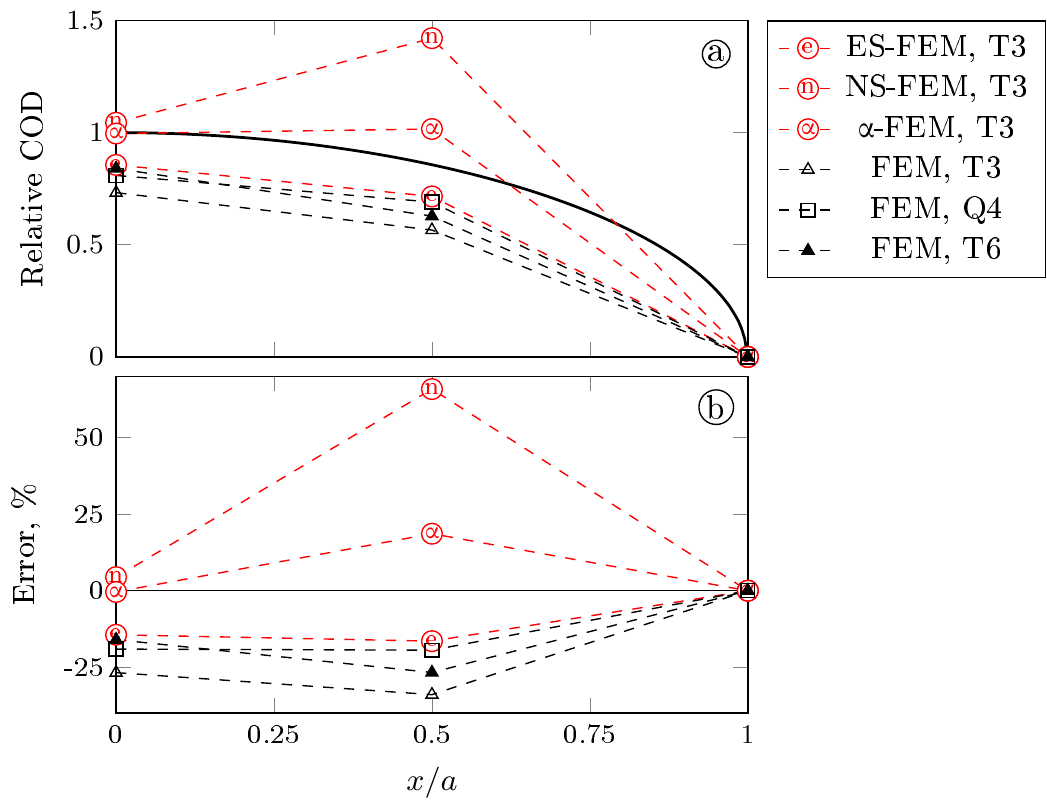}%
    \caption{Comparison of relative crack opening displacements (COD) for the quarter of the CEN specimen (a) and their accuracy (b) calculated using different methods. Mesh density -- two linear or one quadratic elements per half crack length}%
    \label{fig:cen_4_displ_nel4}%
\end{figure}

Qualitatively, the same results for CEN specimen (Fig.~\ref{fig:cen_4_displ_nel4}) look very similar. Only the difference between displacements calculated by traditional FEM using the simplest T3 elements and other results is not as big as previously. Quantitatively, accuracy of all but $\alpha$-FEM methods is twice higher than in the previous case. Again, the most accurate results were obtained by $\alpha$-FEM for $\alpha=0.4$. This value was used for all CEN results presented here.

\begin{figure}%
    \centering
    \includegraphics{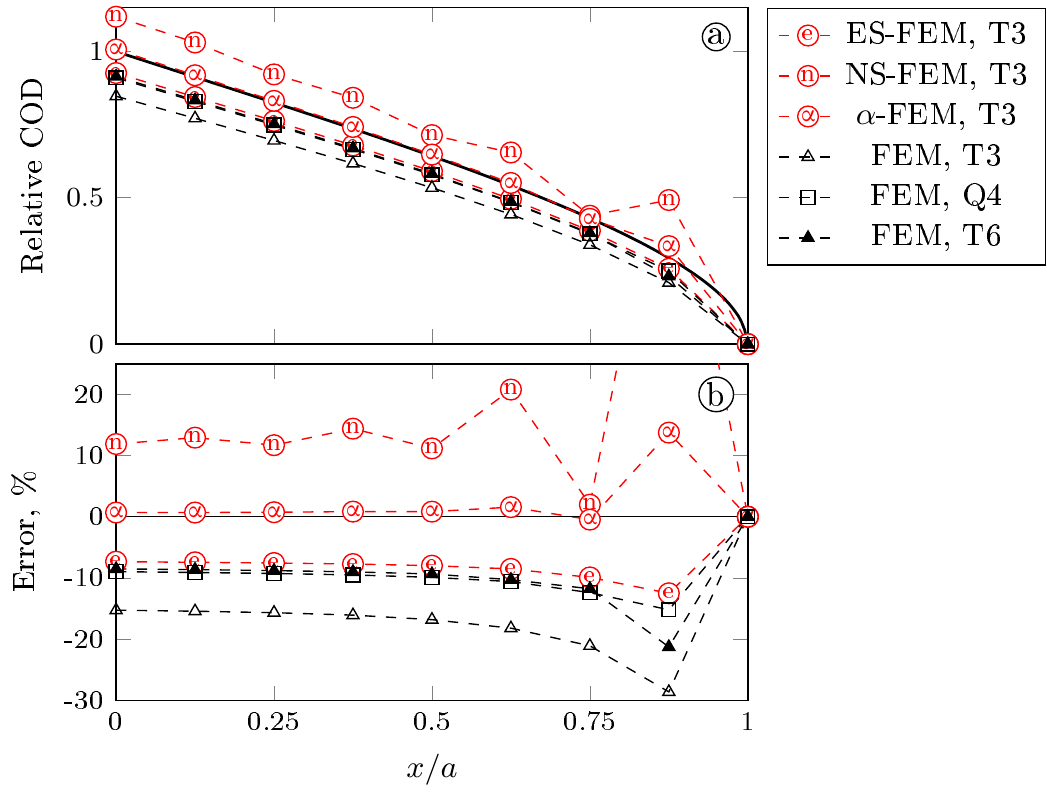}%
    \caption{Comparison of relative COD for the half of the SEN specimen (a) and their accuracy (b) calculated using different methods. Mesh density -- eight linear or four quadratic elements per crack length}%
    \label{fig:sen_4_displ_nel16}%
\end{figure}

For the SEN specimen, increasing number of finite element nodes on the crack surface from 3 to 9 (means decreasing $h/a$ value from 1/2 to 1/8) did not change qualitatively the 'accuracy ranking' of the tested methods (Fig.~\ref{fig:sen_4_displ_nel16}). Again the highest error for calculated COD values was observed for the simplest T3 elements and traditional FEM. Twice more accurate results were obtained using T3 (ES-FEM), Q4 and T6 (both traditional FEM) elements were only a bit less precise. 

NS-FEM overestimates COD values and additionally presents non-physical oscillations of displacements near the crack tip. Our results show that this phenomenon is typical for this type of S-FEM and is independent of mesh density. Of course, this causes similar oscillations of smaller amplitude for $\alpha$-FEM results,  which still remain the most accurate. In fact, displacements for all but first node on the crack surface calculated using this method are practically exact. 

\begin{figure}%
    \centering
    \includegraphics{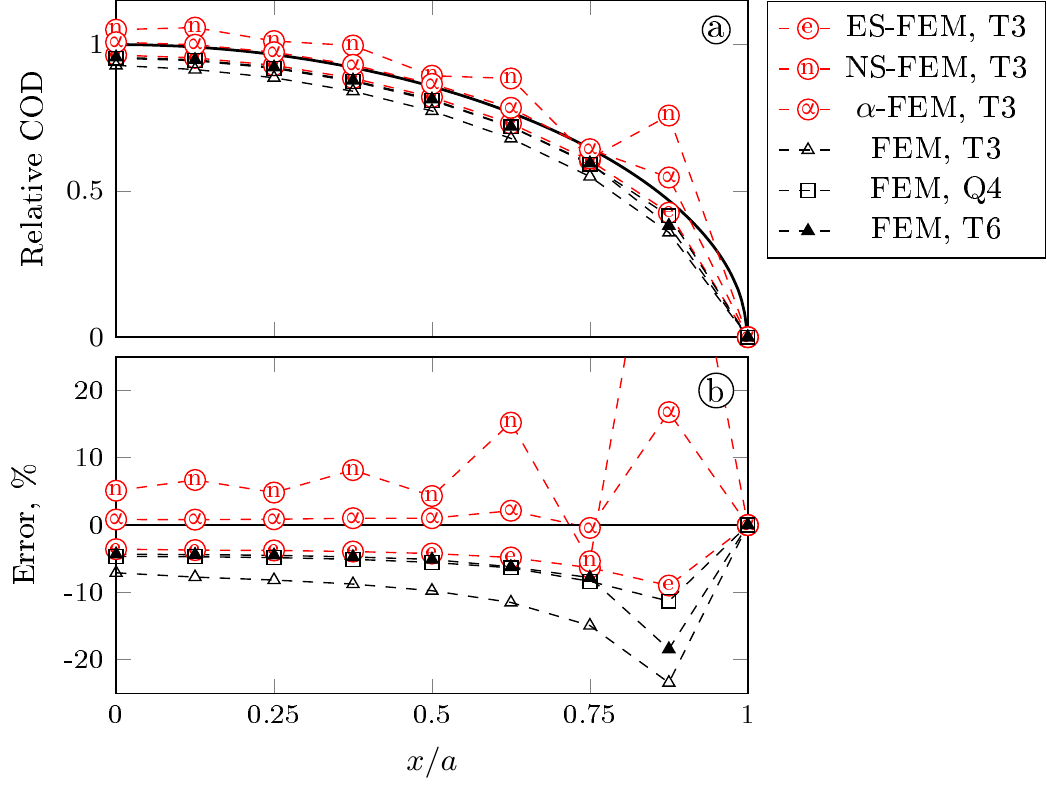}%
    \caption{Comparison of relative COD for the quarter of the CEN specimen (a) and their accuracy (b) calculated using different methods. Mesh density -- eight linear or four quadratic elements per half crack length}%
    \label{fig:cen_4_displ_nel16}%
\end{figure}

Again, for the CEN specimen results are qualitatively similar (Fig.~\ref{fig:cen_4_displ_nel16}). It is worth noting exceptionally high oscillations of the NS-FEM calculated displacements near the crack tip. Due to these oscillations, COD in the second node underestimates accurate value contrary to the general tendency of NS-FEM.

\begin{figure}%
    \centering
    \includegraphics{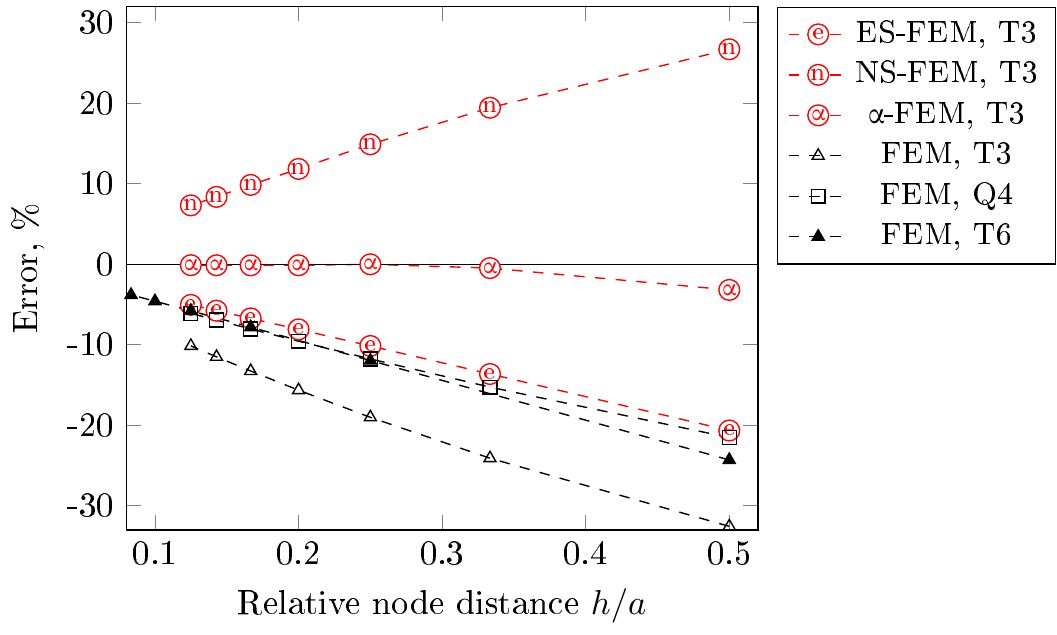}%
    \caption{Accuracy of SIF calculation using different FE elements and methods vs. relative mesh density for SEN specimen}%
    \label{fig:mcci_sen}%
\end{figure}

Accuracy of SIF values calculated by MCCI method for different meshes for SEN specimen (Fig.~\ref{fig:mcci_sen}) follows the trend shown in previous figures. For each mesh density, the worst results were obtained using T3 elements with traditional FEM and NS-FEM. Much more accurate data gave ES-FEM using T3 elements and traditional FEM using Q4 and T6 elements. SIF values calculated by $\alpha$-FEM are the most accurate. Even for the most coarse mesh with $h/a=1/2$ error is about 3\%, for more fine meshes results are practically exact.

\begin{figure}%
    \centering
    \includegraphics{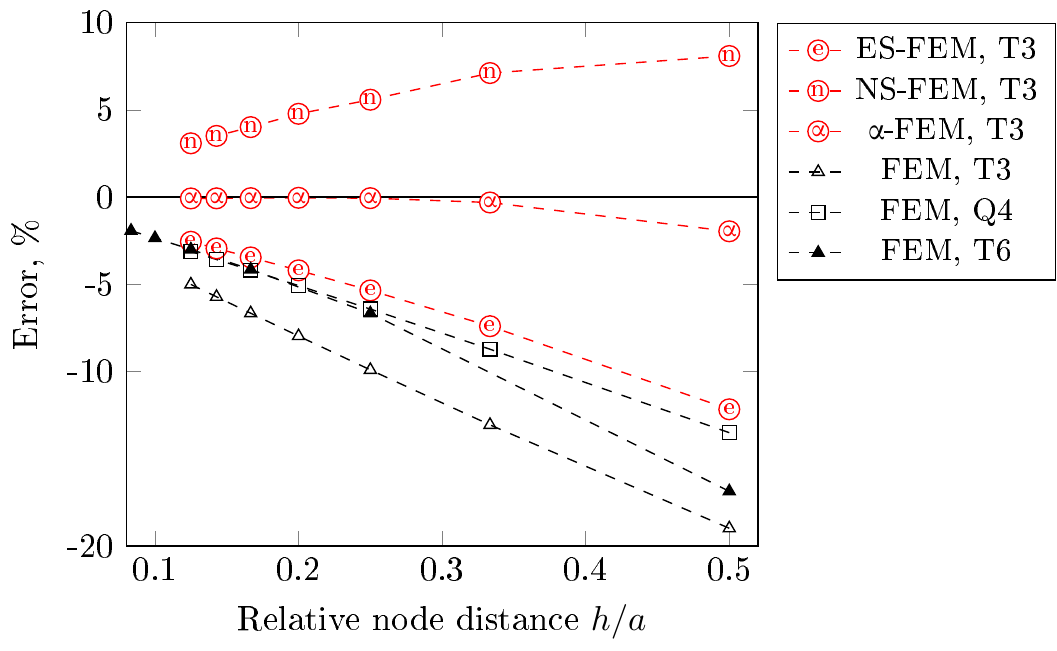}%
    \caption{Accuracy of SIF calculation using different FE elements and methods vs. relative mesh density for CEN specimen}%
    \label{fig:mcci_cen}%
\end{figure}

For each mesh density considered, about twice more accurate SIF values were obtained for CEN specimen compared to the SEN one (Fig.~\ref{fig:mcci_cen})). Superiority of ES-FEM results over the traditional FEM for Q4 and T6 elements is a bit more noticeable here. Again, SIF values calculated by $\alpha$-FEM are the exceptionally accurate compared with all other methods.

It is easy to note almost linear convergence rate of all but $\alpha$-FEM SIFs with respect to $h/a$ for both considered problems. Thus, simple linear extrapolation can be used to improve SIF accuracy if the results for two different mesh densities are known. Even for the most coarse meshes (with $h/a=1/2$ and $h/a=1/3$) this procedure allows obtaining SIF values with accuracy about a few percents.

\section{Conclusions}

The most spectacular results in this study were obtained using $\alpha$-FEM. This method allowed obtaining SIF values with high accuracy even on the most coarse mesh using the simplest linear T3 elements. It is difficult to overestimate practical importance of this fact. 

\begin{figure}%
    \centering
    \includegraphics[width=\textwidth]{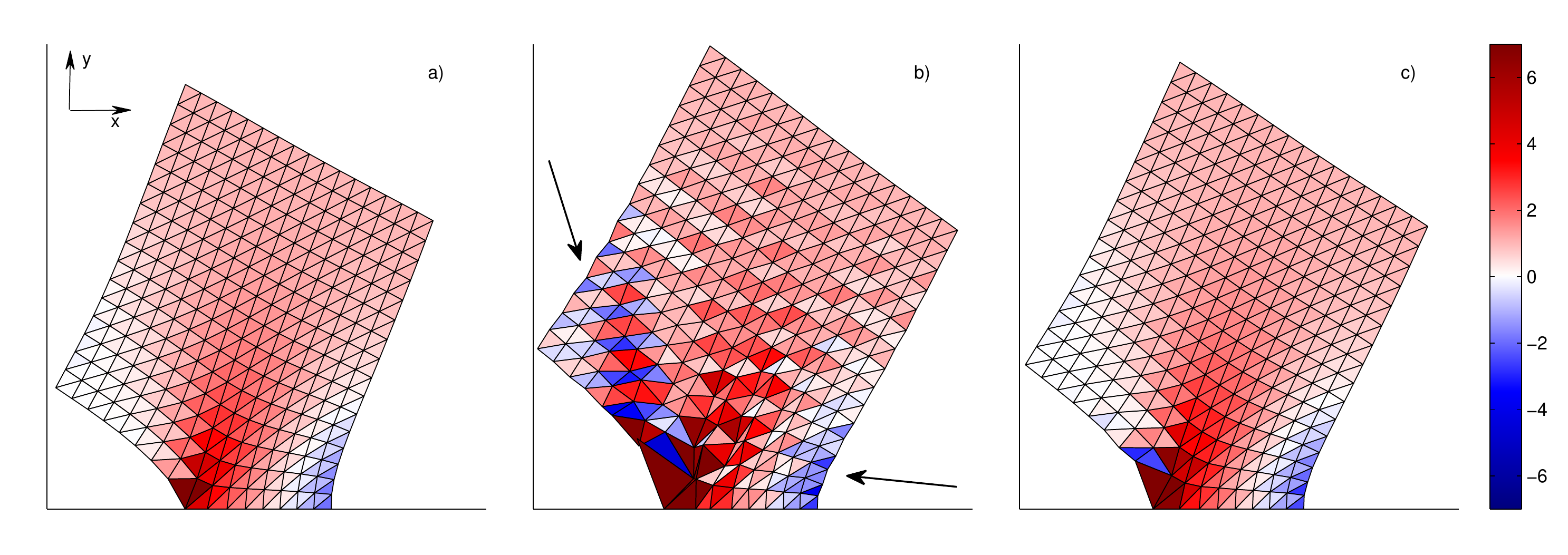}%
    \caption{Distribution of $\sigma_{yy}$ normalized by applied load for the half of the SEN specimen calculated by traditional FEM (a), NS-FEM (b) and $\alpha$-FEM (c). In all subfigures deformation of the mesh is exaggerated using the same scale factor.}%
    \label{fig:sen_stress_yy_nel16}%
\end{figure}

On the other hand, $\alpha$-FEM is a result of combining of the traditional FEM and NS-FEM. The latter method results has a feature that throws into question accurateness of the gradient smoothing technique used in NS-FEM. Calculations performed in this study showed existence of essential oscillations of the displacements determined by NS-FEM near the singularity point. These oscillations are clearly visible in Fig.~\ref{fig:sen_stress_yy_nel16}b as distortions of the element shapes in a broad zone around the crack tip. These distortions cause wavy shape of parts of the specimen borders (marked by arrows in Fig.~\ref{fig:sen_stress_yy_nel16}b).

\begin{figure}%
    \centering
    \includegraphics[width=0.7\textwidth]{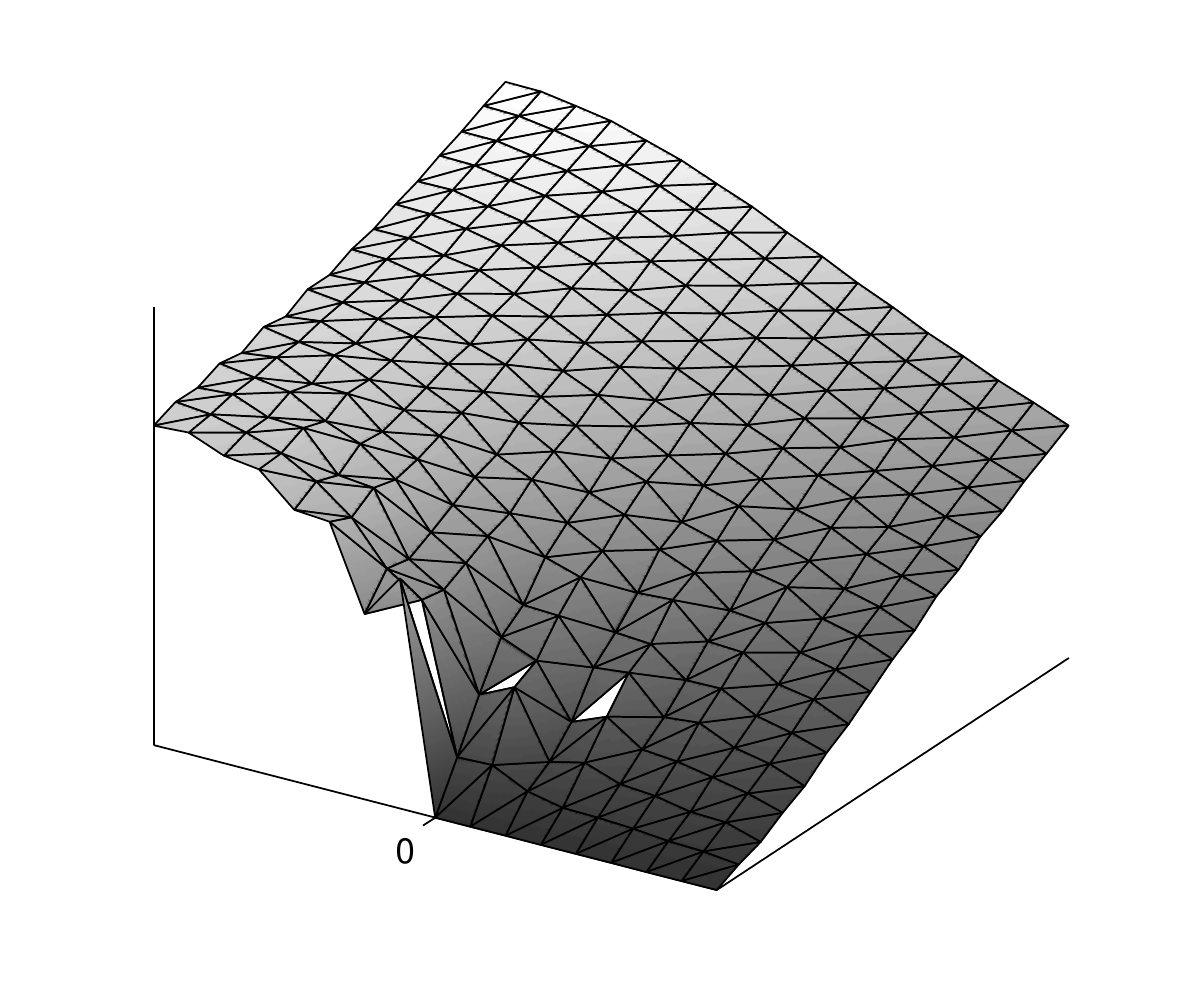}%
    \caption{Displacements normal to the crack face for the quarter of the CEN specimen calculated by NS-FEM (exaggerated). Crack tip position is marked by 0.}%
    \label{fig:cen_displ_yy_nel16}%
\end{figure}

In a bit different graphics form, similar oscillations are shown in Fig.~\ref{fig:cen_displ_yy_nel16} for CEN specimen. To verify all these results and eliminate a possibility that oscillations are caused by a program bug, all computations were repeated using computer code created by the S-FEM authors which is available freely at the companion website of their book~\cite{Liu2010}. Results of calculations were exactly the same as presented here.

Of course, displacements oscillations lead to even more noticeable unphysical oscillations of strain and stress both in the case of NS-FEM (Fig.~\ref{fig:sen_stress_yy_nel16}b) and $\alpha$-FEM (Fig.~\ref{fig:sen_stress_yy_nel16}c). Increasing mesh density causes decreasing of magnitude of oscillations for displacements but not for stresses. It means, that for the problems considered in this study, contrary to its name, NS-FEM does not improve smoothness of the results compared with traditional FEM. The same is true for the $\alpha$-FEM. 

In both cases potential energy of the FE model changes smoothly and has some useful for SIF determination features. However, unphysical behavior of the nodal and elemental results limits reliability of both methods at least for linear fracture mechanics problems.

Finding of NS-FEM instability near singularity points was the main reason to limit the number of cases considered in this study (only two types of geometry with one relative crack length). Further research in this direction has sense only after elimination of the NS-FEM drawbacks.

Accuracy of SIFs calculated using ES-FEM is quite modest when compared with $\alpha$-FEM. In this case, however, there were no problems with stability of displacements and stresses, the method itself is simple, computationally efficient and robust. Results obtained using ES-FEM had the same or even better accuracy compared with T6 elements in traditional FEM. It makes ES-FEM a potentially attractive substitution for traditional FEM in various simple FEA-in-CAD programs like SOLIDWORKS Simulation where quadratic T-elements are widely used.

\bibliographystyle{elsarticle-num}
\bibliography{sif_sfem}

\begin{thebibliography}{10}
\expandafter\ifx\csname url\endcsname\relax
  \def\url#1{\texttt{#1}}\fi
\expandafter\ifx\csname urlprefix\endcsname\relax\def\urlprefix{URL }\fi
\expandafter\ifx\csname href\endcsname\relax
  \def\href#1#2{#2} \def\path#1{#1}\fi

\bibitem{Liu2010}
G.~Liu, T.~{Nguyen-Thoi}, Smoothed Finite Element Methods, CRC Press, 2010.

\bibitem{Liu2011}
G.R.~Liu, N.~Nourbakhshnia, Y.~Zhang, A novel singular {ES-FEM} method for simulating
  singular stress fields near the crack tips for linear fracture problems,
  Eng. Fract. Mech. 78 (2011) 863--876.
\newblock \href {http://dx.doi.org/10.1016/j.engfracmech.2009.11.004}
  {\path{doi:10.1016/j.engfracmech.2009.11.004}}.

\bibitem{Nguyen2012}
H.~Nguyen-Xuan, G.R.~Liu, N. {Nourbakhshnia}, L.~Chen, A novel singular {ES-FEM} for crack
  growth simulation, Eng. Fract. Mech. 84 (2012) 41--66.
\newblock \href {http://dx.doi.org/10.1016/j.engfracmech.2012.01.001}
  {\path{doi:10.1016/j.engfracmech.2012.01.001}}.

\bibitem{Zeng2013}
W.~Zeng, G.R.~Liu, Y. {Kitamura}, H.~{Nguyen-Xuan}, A three-dimensional {ES-FEM} for
  fracture mechanics problems in elastic solids, Eng. Fract. Mech.
  114 (2013) 127--150.
\newblock \href {http://dx.doi.org/10.1016/j.engfracmech.2013.10.017}
  {\path{doi:10.1016/j.engfracmech.2013.10.017}}.

\bibitem{barsoum1975}
R.~Barsoum, Further application of quadratic isoparametric finite elements to
  linear fracture mechanics of plate bending and general shells, Int. J. Fracture 11 (1975) 167--169.
\newblock \href {http://dx.doi.org/10.1007/BF00034724}
  {\path{doi:10.1007/BF00034724}}.

\bibitem{Fu2012}
P.~Fu, S.M.~Johnson, R.R. Settgast, C.~Carrigan, Generalized displacement correlation
  method for estimating stress intensity factors, Eng. Fract. Mech. 88 (2012) 90--107.
\newblock \href {http://dx.doi.org/10.1016/j.engfracmech.2012.04.010}
  {\path{doi:10.1016/j.engfracmech.2012.04.010}}.

\bibitem{es}
G.R. Liu, T. {Nguyen-Thoi}, K.Y. Lam, An edge-based smoothed finite element method ({ES-FEM})
  for static, free and forced vibration analyses in solids, J. Sound Vib. 320
  (2009) 1100--1130.
\newblock \href {http://dx.doi.org/110.1016/j.jsv.2008.08.027}
  {\path{doi:110.1016/j.jsv.2008.08.027}}.

\bibitem{ns}
G.R. Liu, T. {Nguyen-Thoi}, H. {Nguyen-Xuan}, K.Y. Lam, A node-based smoothed finite
element method ({NS-FEM}) for upper bound solutions to solid mechanics problems,
  Comput. Struct. 87 (2009) 14--26.
\newblock \href {http://dx.doi.org/10.1016/j.compstruc.2008.09.003}
  {\path{doi:10.1016/j.compstruc.2008.09.003}}.

\bibitem{alpha}
G.R. Liu, T. {Nguyen-Thoi},  K.Y. Lam, A novel alpha finite element method
  ($\alpha${FEM}) for exact solution to mechanics problems using triangular and
  tetrahedral elements, Comput. Method. Appl. M.
  197 (2008) 3883--3897.
\newblock \href {http://dx.doi.org/doi:10.1016/j.cma.2008.03.011}
  {\path{doi:doi:10.1016/j.cma.2008.03.011}}.

\bibitem{adina}
ADINA R\&D Inc., Watertown, ADINA Theory and Modeling Guide (2013).

\bibitem{frac}
A.~Ingraffea, I.~Wawrzynek, Finite
  Element Methods for Linear Elastic Fracture Mechanics. In: Comprehensive Structural Integrity, Ed.:I.~Milne, R.O.~Ritchi and B.~Karihaloo, Elsevier, vol. 3 (2003) 1--88.

\bibitem{mcci}
R.~Krueger, Virtual crack closure technique: History, approach, and
  applications, Appl. Mech. Rev. 57 (2004) 109--143.
\newblock \href {http://dx.doi.org/10.1115/1.1595677}
  {\path{doi:10.1115/1.1595677}}.

\end{thebibliography}

\end{document}